\documentclass[11pt,a4paper]{article}

\usepackage[a4paper,margin=1in]{geometry}
\usepackage[T1]{fontenc}
\usepackage[utf8]{inputenc}
\usepackage{amsmath}
\usepackage{newtxtext,newtxmath}
\usepackage{textcomp}
\usepackage{microtype}
\usepackage{graphicx}
\usepackage{float}
\usepackage{booktabs}
\usepackage{caption}
\usepackage{indentfirst}
\usepackage{xurl}
\usepackage{newunicodechar}
\usepackage[
  backend=biber,
  style=numeric-comp,
  sorting=none,
  sortcites=true,
  giveninits=false,
  maxbibnames=99,
  minbibnames=99,
  doi=true,
  url=true,
  isbn=false
]{biblatex}
\usepackage[hidelinks]{hyperref}

\graphicspath{{figures/}}
\DeclareNameAlias{default}{given-family}
\DeclareFieldFormat{doi}{\url{https://doi.org/#1}}
\DeclareFieldFormat{url}{\url{#1}}

\makeatletter
\renewbibmacro*{cite:dump}{%
  \usebibmacro{cite:dump:inset}%
  \ifnumgreater{\value{cbx@tempcnta}}{0}
    {\setunit{\multiciterangedelim}%
     \usebibmacro{cite:print:last:labelnumber}%
     \global\undef\cbx@lastprefix}
    {}%
  \setcounter{cbx@tempcnta}{0}}
\makeatother

\newunicodechar{α}{\ensuremath{\alpha}}
\newunicodechar{β}{\ensuremath{\beta}}
\newunicodechar{γ}{\ensuremath{\gamma}}
\newunicodechar{θ}{\ensuremath{\theta}}
\newunicodechar{λ}{\ensuremath{\lambda}}
\newunicodechar{≈}{\ensuremath{\approx}}
\newunicodechar{₀}{\textsubscript{0}}
\newunicodechar{₁}{\textsubscript{1}}
\newunicodechar{₂}{\textsubscript{2}}
\newunicodechar{₃}{\textsubscript{3}}
\newunicodechar{₅}{\textsubscript{5}}

\begin{document}

\begin{center}
{\LARGE\bfseries MatDiffract: A Material-Informed Automated Analysis Platform for X-ray Powder Diffraction\par}
\vspace{1.2em}
{\large Hongqing Wang\textsuperscript{123}, Mingwei Chen\textsuperscript{45}, Hongjie Luo\textsuperscript{126}, Wen Yin\textsuperscript{12}, Fazhi Qi\textsuperscript{12}, Xuqing Chai\textsuperscript{3}*, Miao Liu\textsuperscript{45}*, Fengyao HOU\textsuperscript{12}*\par}
\vspace{0.8em}
{\small \textsuperscript{1}Institute of High Energy Physics, Chinese Academy of Sciences, Beijing, China\\[2pt]
\textsuperscript{2}Spallation Neutron Source Science Center, Dongguan, China\\[2pt]
\textsuperscript{3}College of Computer and Information Engineering, Henan Normal University, Xinxiang, China\\[2pt]
\textsuperscript{4}Dongguan Institute of Materials Science and Technology, Chinese Academy of Sciences, Dongguan, China\\[2pt]
\textsuperscript{5}Songshan Lake Material Laboratory, Dongguan, China\\[2pt]
\textsuperscript{6}School of Computer Science, Sun Yat-sen University, Guangzhou, China\par}
\end{center}
\vspace{1em}

\begin{abstract}

High-throughput experimentation and self-driving laboratories are drastically accelerating materials discovery, yet automated interpretation of X-ray powder diffraction (XRPD) data remains a critical rate-limiting step. Conventional search-match workflows rely heavily on expert manual intervention, while pure data-driven machine learning approaches suffer from limited generalizability across chemical systems and lack rigorous crystallographic interpretability. Here we present MatDiffract, a material-informed automated analysis platform for high-throughput XRPD characterization.

Built on a first-principles density functional theory (DFT)-derived inorganic crystal structure database, Atomly, MatDiffract constructs a perturbation-augmented simulated diffraction database, embeds multi-scale diffraction features into indexable vectors, and integrates hierarchical vector retrieval with full-pattern fitting Rietveld refinement and quantitative phase fitting. Benchmarked on 875 single-phase experimental patterns, the platform achieves 91.3\% Top-1 and 97.2\% Top-10 identification accuracy after automated refinement. For binary and ternary multiphase mixtures, it delivers 85.0\% and 70.0\% Top-1 accuracy with mass fraction mean absolute errors as low as 1.2\% and 1.8\%, respectively. Beyond mere phase labeling, MatDiffract outputs full crystallographic results including refined structural models, fitted profiles, and quantitative compositions within tens of seconds per sample. Its modular vector-based architecture supports seamless incremental expansion to new material systems, providing an end-to-end solution to close the characterization throughput gap for autonomous materials discovery and high-throughput materials development.

\end{abstract}

\noindent\textbf{Keywords: X-ray powder diffraction (XRPD), vector retrieval, phase identification, Rietveld refinement}

\section{Introduction}\label{sec:introduction}

The advent of high-throughput computation, automated synthesis platforms, and self-driving laboratories has transformed the pace of materials discovery, enabling systematic screening of vast chemical spaces at unprecedented scale\supercite{szymanski2023,yotsumoto2024,yu2025,maruyama2020}. X-ray powder diffraction (XRPD) remains the most widely deployed characterization technique for verifying phase purity, determining crystal structures, and quantifying phase assemblages across nearly all classes of crystalline materials\supercite{hanawalt1938,frevel1976,rietveld1969,hill1987,ali2022,lopresti2022}. However, the throughput of XRPD data interpretation has failed to keep pace with advances in automated synthesis and data acquisition. Conventional analysis workflows depend on time-consuming manual inspection and expert judgment to resolve peak overlap, distinguish competing phase candidates, and perform structural refinement. This human-in-the-loop bottleneck severely limits the throughput of closed-loop materials discovery workflows, creating an urgent demand for robust, fully automated XRPD interpretation tools.

Existing efforts to automate XRPD analysis fall broadly into two categories: conventional reference-based search-match methods and emerging data-driven machine learning approaches. Standard search-match protocols retrieve candidate phases from curated experimental databases but require extensive manual refinement to validate candidates and quantify compositions; performance degrades sharply for multiphase samples with severe peak overlap and low-abundance minor phases\supercite{ali2022,lopresti2022}. Recent advances in deep learning, generative models, and contrastive learning have shown promise for accelerated phase identification, but these methods typically rely on system-specific training data, require full retraining when extended to new chemical systems, and often produce phase labels without accompanying crystallographic interpretability or structural validation\supercite{suzuki2022,salgado2023,davel2025,le2023,zhang2024,chen2024,lee2020,mathew2026,xing2025,cao2025,oviedo2019,vecsei2019,zaloga2020,greasley2023,yu2026,xu2026}. Moreover, label-only predictive models fail to reliably discriminate between competing candidates and provide limited crystallographic insight, since individual PXRD patterns are inherently compatible with multiple phase assemblages\supercite{parackal2024,fei2026}. To date, no unified framework has combined high identification accuracy, rigorous physical interpretability, native support for multiphase quantification, and seamless extensibility to new material systems, hindering the integration of automated diffraction analysis into high-throughput discovery workflows.

In this work, we introduce MatDiffract, a material-informed automated PXRD analysis platform designed to address this gap, built upon the first-principles inorganic crystal structure database, Atomly\supercite{liu2023,xie2024}. MatDiffract adopts a physics-informed vector retrieval architecture: it first generates a comprehensive database of perturbation-augmented simulated PXRD patterns from DFT-optimized crystal structures, embeds multi-dimensional diffraction features into a searchable vector database, and performs hierarchical vector retrieval to rapidly identify candidate phase assemblages. The retrieved candidates are then validated and ranked via full-pattern fitting, Rietveld refinement, and mass fraction optimization, delivering quantitatively interpretable crystallographic results rather than mere classification labels. We benchmark the platform on both single-phase and multiphase experimental PXRD datasets, demonstrating a higher identification accuracy and quantitative precision. We further show that the modular vector-based architecture supports incremental expansion to new crystal structures without model retraining, making the framework broadly applicable to high-throughput materials screening and autonomous experimental workflows.

\section{Architecture and Workflow}\label{sec:architecture}

\begin{figure}[H]
\centering
\includegraphics[width=\linewidth,height=0.78\textheight,keepaspectratio]{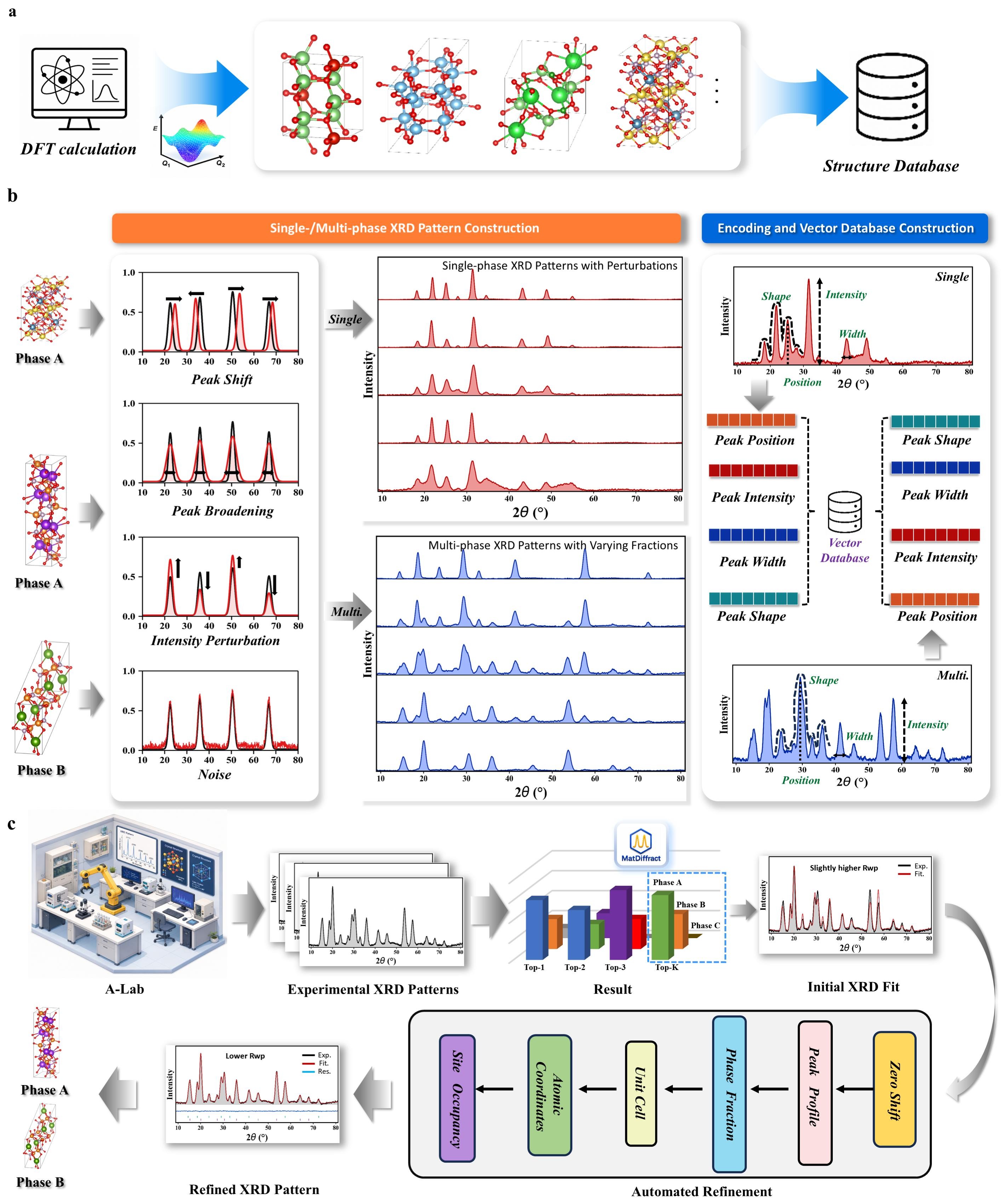}
\caption{Schematic overview of the MatDiffract workflow. (a) DFT-optimized crystal structures are stored with structural metadata. (b) Simulated XRPD patterns are generated with controlled perturbations to peak shape, intensity, background and noise, paired with synthetic multiphase patterns with predefined phase combinations and phase fractions. Peak positions, intensities, widths, and local profile features are extracted and embedded into the vector database. (c) Experimental XRPD patterns are mapped into the same feature space. Hierarchical vector retrieval returns ranked candidates for subsequent full-pattern fitting and structural refinement.}
\label{fig:workflow}
\end{figure}

The MatDiffract workflow comprises four stages (Figure~\ref{fig:workflow}). First, a simulated XRPD pattern database is built based on crystal structures from the Atomly database, augmented with peak shifts, broadening, intensity variations, and background noise to mimic common experimental artifacts. Second, the peak features of positions, intensities, widths, and local profile are extracted and stored in a vector database paired with the corresponding structural, perturbation and compositional metadata. Third, experimental patterns are embedded into the same feature vector space and retrieved via hierarchical vector search to return single-phase or multiphase candidates. Fourth, the candidates are refined by full-pattern fitting, structural refinement, and mass-fraction optimization to output the final phase composition, residual profile, mass fractions, and structural parameters.

\subsection{Simulated XRPD Pattern Database Construction}\label{sec:simulated-database}

After curating the DFT-derived inorganic crystal structure database to exclude redundant and thermodynamically unstable entries\supercite{liu2023}, XRPD patterns were simulated for Cu Kα radiation (λ = 1.5418 Å). Experimental patterns deviate from ideal simulations due to instrumental offsets, microstructure effects, and measurement conditions, with solid solution, microstrain, or defects inducing peak shift, broadening and intensity redistribution\supercite{riaz2023,madisetti2025}. Each single-phase simulated pattern was therefore augmented via controlled perturbation of peak positions, widths, intensities, and background noise, subject to crystallographic constraints. Multiphase simulated patterns were then generated from predefined phase combinations and mass fractions.

\subsection{XRPD Vector Database Construction}\label{sec:vector-database}

To accelerate retrieval, MatDiffract recasts phase identification as a high-dimensional similarity search task using vector database infrastructure and approximate nearest-neighbor algorithms\supercite{malkov2020,johnson2021,wang2021,pan2024,zhang2023}. The peak features of position, intensity, width, and local-profile are extracted from each simulated pattern and stored as indexable vectors. Local profile features complement discrete peak parameters, distinguishing structures with similar peak locations but different shoulders, weak peaks, or line shapes. Unlike closed-label classifiers, the vector database supports incremental expansion without redefining labels or retraining and preserves complete metadata for traceability and downstream refinement.

\subsection{Feature Recognition and Vector Retrieval}\label{sec:retrieval}

Preprocessed experimental XRPD patterns are featured and queried via a hierarchical retrieval strategy that sequentially applies the features of peak positions, intensities, and local profile to balance recall and efficiency. Peak positions, governed by lattice parameters, are relatively robust to intensity variations. Peak intensities are more sensitive to preferred orientation, morphology, and instrumental effects. Local profile shape and peak width encode fine structural details. The retrieval stage returns a ranked Top-k candidate set, not the final assignment, providing a high-recall pool for subsequent full-pattern fitting and structural refinement.

\subsection{Full-Pattern Refinement}\label{sec:refinement}

Retrieved candidates are refined against the experimental pattern via a two-step procedure: global peak position offset correction, followed by optimization of profile shape, background, and scale factors. For multiphase samples, the mass fractions of each phase are optimized to match the total diffraction intensity. For single-phase samples, the lattice parameters and relevant local structural parameters are further refined to match the experimental data. The final output includes phase combination, mass fractions, residual profile, Bragg positions and structural parameters, enabling user evaluation of fit credibility.

\section{Results and Discussion}\label{sec:results}

The performance of MatDiffract was systematically evaluated using two experimentally measured XRPD datasets, comprising single-phase and multiphase samples respectively. The single-phase dataset was constructed to assess phase identification accuracy and automated full-pattern fitting refinement for patterns dominated by a single crystalline phase. It was compiled from the RRUFF mineral database, which provides curated experimental XRPD patterns paired with corresponding structural refinement data\supercite{lafuente2015}. After excluding entries with missing patterns, incomplete structural information, or duplication, 875 valid experimental patterns were retained for benchmarking. The multiphase dataset consists of 40 experimental patterns collected from laboratory-synthesized mixtures to evaluate phase-combination identification and mass-fraction quantification under conditions representative of practical materials characterization, including phase coexistence, severe peak overlap, and low-abundance minor phases. All 40 experimental samples were prepared on the automated A-Lab synthesis platform via a standardized workflow of batching, weighing, grinding, and thermal treatment, with precisely predefined phase compositions. The dataset covers binary and ternary mixtures with varied mass-fraction distributions. XRPD data were collected on a Bruker D8 Discover diffractometer using Cu Kα radiation (λ = 1.5418 Å), over a 2θ range of 10°--90° with a step size of 0.02°. The average analysis time per pattern was also quantified to assess the practical suitability of MatDiffract for high-throughput XRPD workflows.

\subsection{Single-Phase Identification and Automated Refinement}\label{sec:single-phase}

The performance of MatDiffract on the single-phase dataset was evaluated to provide a straightforward benchmark of the platform's ability to identify the target structure from real experimental data. The curated dataset of 875 experimental patterns was used to assess both phase identification retrieval accuracy and the quality of automated Rietveld refinement.

\begin{figure}[H]
\centering
\includegraphics[width=\linewidth,height=0.78\textheight,keepaspectratio]{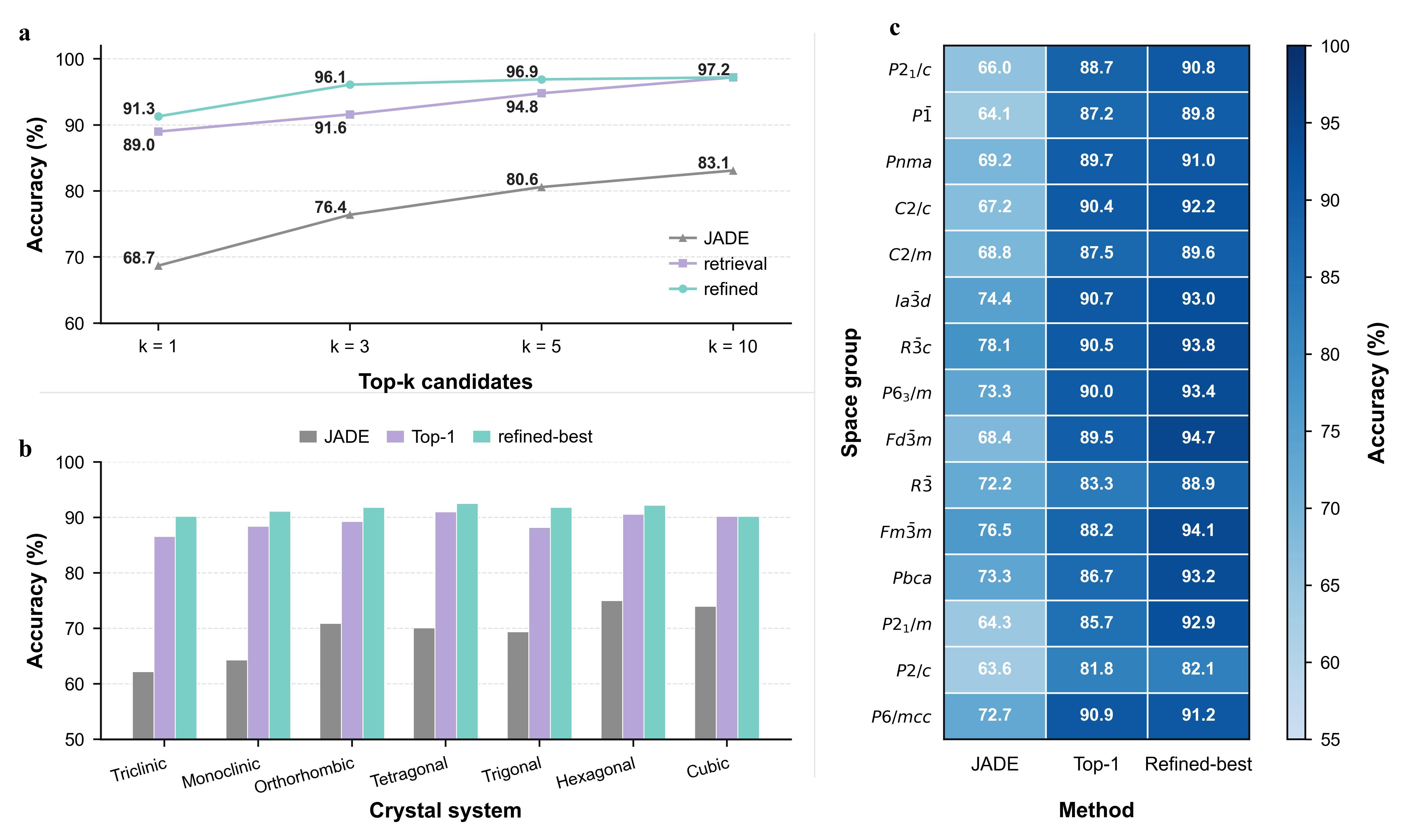}
\caption{Single-phase XRPD phase identification performance. (a) Identification accuracy of JADE, initial MatDiffract retrieval, and the full automated-refinement results, as a function of Top-k cutoff. (b) Top-1 accuracy of JADE, initial MatDiffract retrieval, and the best refined MatDiffract result across all seven crystal systems. (c) Top-1 accuracy heat map for JADE, initial MatDiffract Top-1 retrieval, and the best refined result across the 15 most frequent space groups.}
\label{fig:single-phase}
\end{figure}

Benchmark comparisons were performed against JADE, a widely adopted commercial XRPD analysis suite for phase identification, quantitative phase analysis, and full-pattern fitting. JADE was operated with prior elemental information, such that its reported accuracy corresponds to element-constrained search rather than blind open-composition identification. Figure~\ref{fig:single-phase}a compares the Top-k identification accuracy of JADE, the initial vector retrieval stage of MatDiffract, and the full-pattern fitting stages with automated refinement of MatDiffract. JADE achieved a Top-1 accuracy of 68.7\% and a Top-10 accuracy of 83.1\%. The initial vector retrieval stage of MatDiffract achieved a Top-1 accuracy of 89.0\% and a Top-10 accuracy of 97.2\%. Following automated refinement and reranking, Top-1 accuracy improved to 91.3\%, with Top-3 and Top-5 accuracies reaching 96.1\% and 96.9\%, respectively; Top-10 accuracy remained at 97.2\% because refinement only reranks candidates already retrieved within the initial candidate pool. This indicates that the correct phases are consistently ranked near the top of the candidate list, such that users typically need to inspect only the highest-ranked candidates and the fitted profiles, rather than manually screening long candidate lists. For benchmarking purposes, all Top-k candidates were refined systematically; in fact, the number of refined candidates can be limited to balance accuracy and computational runtime.

Figures~\ref{fig:single-phase}b and \ref{fig:single-phase}c evaluate the robustness of phase identification across crystal systems and space groups. MatDiffract outperformed JADE in all seven crystal systems, with the refined result maintaining an accuracy of about 90\% or higher even for inherently challenging triclinic and monoclinic systems. The space-group heat map shows that the performance gains are not only restricted to high-symmetry structures but also to common low-symmetry groups such as P21/c, P-1, Pnma and C2/c. Since crystal system and space group govern peak distributions, extinction rules, feature counts, and peak overlap, the consistent improvement across all classes stems from robust pattern representations and candidate reranking, rather than favorable bias toward a small subset of simple structures.

This performance aligns with the core design principles of the MatDiffract workflow. Database augmentation via controlled perturbation expands candidate coverage, while multi-feature retrieval integrates peak positions, relative intensities, peak widths, and local profile information, reducing dependence on a small set of intense peaks or simple peak-list matching. The use of perturbed simulated patterns further improves robustness against experimental peak shifts and profile distortions. Most critically, automated refinement constrains the candidate ranking through full-pattern crystallographic interpretability. In XRPD, local peak similarity does not guarantee a physically meaningful full-pattern fit, such that related structures, peak overlap, preferred orientation, and peak-shape variations all produce locally similar but globally inconsistent matches. MatDiffract therefore addresses this by coupling retrieval with automated Rietveld refinement, validating candidate structures using goodness-of-fit metrics (Rwp), residual profiles, and Bragg-position consistency.

\begin{figure}[H]
\centering
\includegraphics[width=\linewidth,height=0.78\textheight,keepaspectratio]{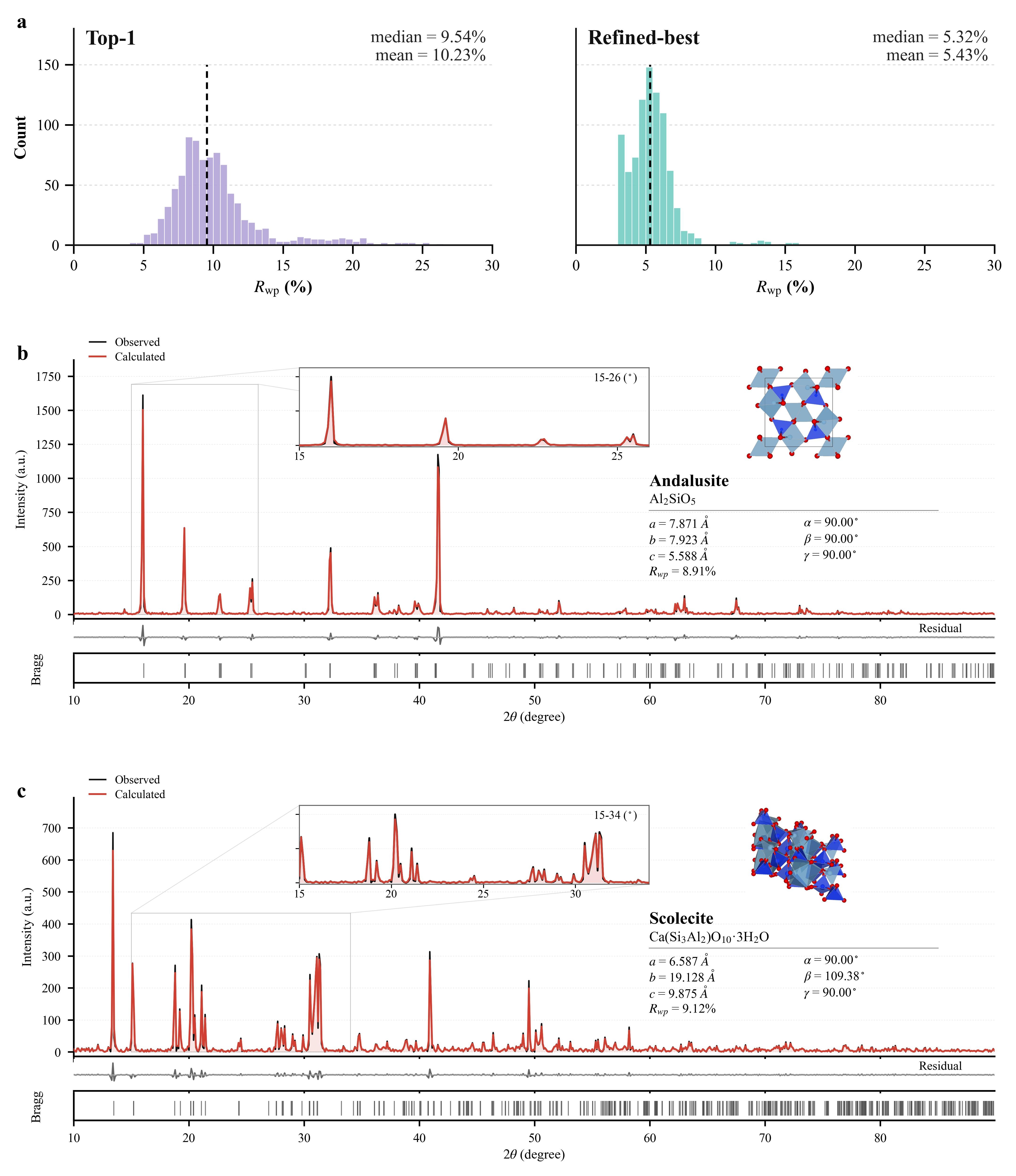}
\caption{Performance of automated Rietveld refinement. (a) Distributions of weighted-profile R-factor (Rwp) for the initial Top-1 retrieved candidate and the best refined candidate; dashed lines mark median values. (b) Representative Rietveld refinement for andalusite (Al₂SiO₅), showing the full diffraction profile, magnified low-angle region, refined lattice parameters, crystal structural model, and Rwp = 8.91\%. (c) Representative refinement for scolecite (Ca(Si₃Al₂)O₁₀·3H₂O), showing the full diffraction profile, magnified low-angle region, refined lattice parameters, crystal structural model, and Rwp = 9.12\%. Black, red, and gray lines represent the experimental pattern, calculated pattern, and fitting residual, respectively; vertical tick marks denote Bragg positions.}
\label{fig:rietveld}
\end{figure}

Figure~\ref{fig:rietveld} characterizes the performance of automated Rietveld refinement from the perspective of full-pattern fitting. Whereas initial retrieval mainly ranks candidates based on the similarity of peak positions, peak intensities, and local profile features, Rietveld refinement optimizes background, lattice parameters, profile shape, and scale factors for a given structural model to minimize discrepancy between calculated and experimental patterns over the full 2θ range. Accordingly, refinement mitigates mismatches arising from experimental artifacts, lattice-parameter deviations, or profile differences and provides a rigorous test of whether a candidate can plausibly account for the complete diffraction pattern.

Figure~\ref{fig:rietveld}a demonstrates that refinement substantially improves candidate validation and reranking. The Top-1 candidates at the initial stage exhibited a broad distribution of Rwp values, with a median of 9.54\% and a mean of 10.23\%, indicating that some highly ranked candidates remained poor full-pattern fits. After automated refinement and reranking, the distribution of best-candidate Rwp shifted markedly toward smaller values, with a median of 5.32\% and a mean of 5.43\%. This confirms that refinement leverages information from the full diffraction profile to distinguish physically plausible structures from locally similar but globally inadequate alternatives.

Figure~\ref{fig:rietveld}b presents a representative refinement for andalusite (Al₂SiO₅). The refined lattice parameters were a=7.871Å, b=7.923Å, c=5.588Å, and α=β=γ=90.00°, with Rwp=8.91\%. The calculated pattern accurately reproduced the positions and relative intensities of all major peaks over the full profile, and the inset covering 15°-26° further shows that both the intense peak profile and nearby weak peaks were well constrained. Refinement therefore elevates a mere phase label into inspectable crystallographic evidence, including lattice parameters, fitted profile, residuals, and Bragg peak positions.

Figure~\ref{fig:rietveld}c shows a representative refinement for scolecite, a low-symmetry hydrated silicate characterized by dense peak packing and substantial local overlap. The refined lattice parameters were a=6.587Å, b=19.128Å, c=9.875Å, α=90.00°, β=109.38°, and γ=90.00°, with Rwp=9.12\%. Despite the abundance of weak and adjacent peaks, the calculated pattern closely follows the main peak groups and overall profile, with residuals concentrated mainly around a few strong or overlapping peaks. This example demonstrates that the workflow reliably validates complex low-symmetry structures as well as simpler high-symmetry patterns.

Collectively, the statistical results in Figure~\ref{fig:rietveld}a and the representative case studies in Figures~\ref{fig:rietveld}b and \ref{fig:rietveld}c illustrate that automated refinement converts raw candidate similarity into quantitative full-pattern interpretability. For candidates with comparable local similarity, full-pattern fitting exploits weak peaks, profile shape and residual information that are not fully constrained during the vector retrieval stage. For already well-identified candidates, refinement delivers physically meaningful lattice parameters and fit quality that enable direct assessment of result credibility. This closed-loop architecture, from rapid retrieval to rigorous structural validation, constitutes a key advantage of MatDiffract over label-only classification or peak-list matching.

\subsection{Multiphase Identification and Quantitative Phase Analysis}\label{sec:multiphase}

Under realistic experimental conditions, the relationship between an XRPD pattern and the underlying phase composition is rarely one-to-one. In multiphase systems, multiple phases contribute diffracted intensity simultaneously, and peak overlap often renders individual Bragg reflections difficult to resolve and assign. Peak broadening, background noise, and finite instrumental resolution further obscure weak diffraction features, such that low-abundance minor phases are readily masked by intense peaks from the major phase. Reliable multiphase XRPD analysis therefore requires both accurate identification of the phase combination that accounts for the full diffraction pattern and quantitative determination of the mass fraction of each constituent phase.

MatDiffract implements a two-stage workflow for multiphase analysis. In the first stage, candidate phase combinations and initial mass-fraction estimates are returned via vector retrieval. In the second stage, these candidates are refined by multiphase full-pattern fitting to obtain the final phase combination and quantitative mass fractions. Accordingly, multiphase performance was evaluated with respect to both phase-combination identification accuracy and quantitative phase analysis precision.

\begin{figure}[H]
\centering
\includegraphics[width=\linewidth,height=0.78\textheight,keepaspectratio]{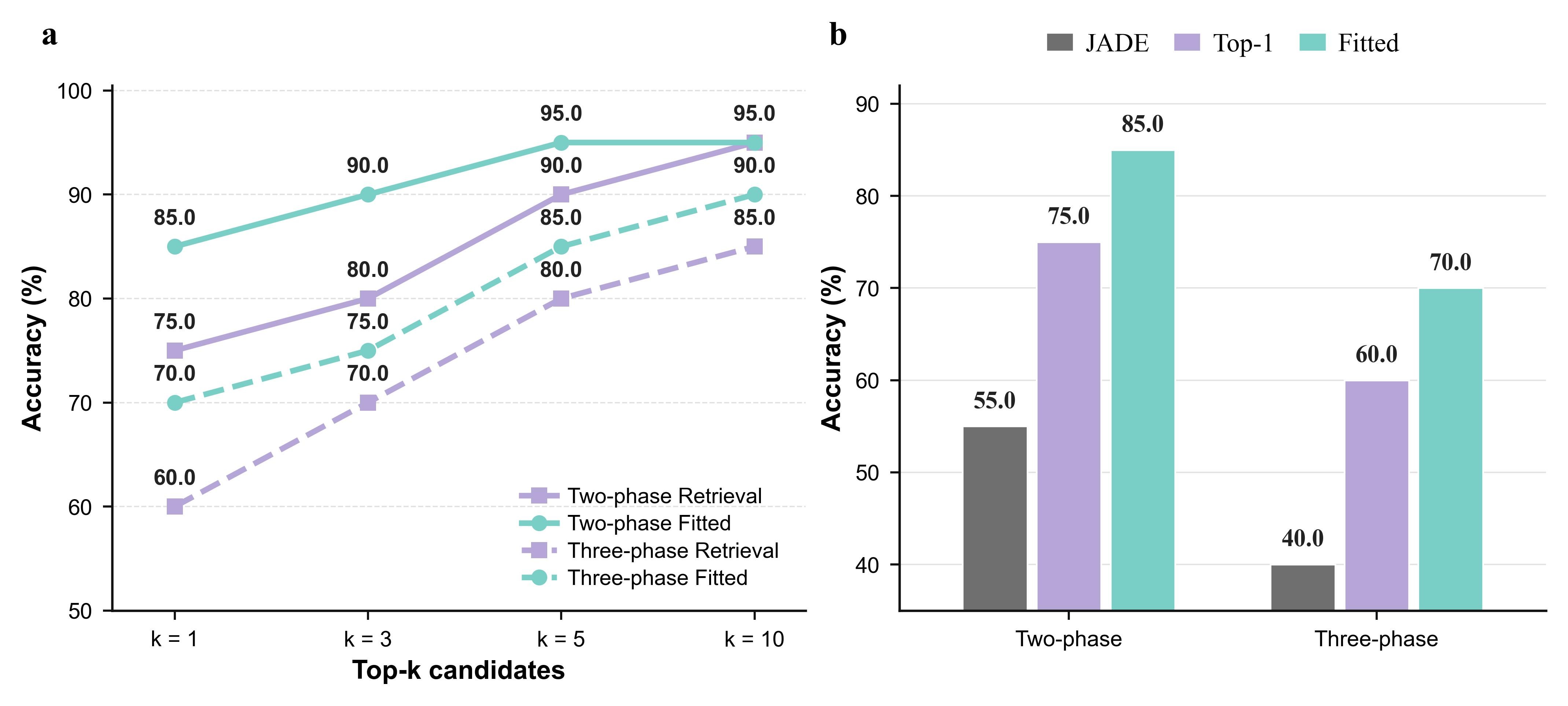}
\caption{Phase-combination identification in multiphase experimental XRPD patterns. (a) Identification accuracy of initial retrieval and full-pattern fitting results for binary and ternary samples as a function of Top-k cutoff. (b) Top-1 identification accuracy comparison of JADE, initial MatDiffract retrieval, and full-pattern fitting MatDiffract results.}
\label{fig:multiphase-identification}
\end{figure}

To evaluate MatDiffract under conditions representative of routine materials characterization, binary and ternary powder mixtures were prepared from well-characterized phase powders at predefined mass ratios. Nominal compositions serve only as an initial reference, because minor phase changes, contamination, or oxidation can occur during preparation, storage, or measurement. Manually refined Rietveld results were therefore adopted as the ground-truth benchmark. All multiphase patterns were manually refined using GSAS-II\supercite{toby2013} by researchers with extensive expertise in XRPD analysis and Rietveld refinement, considering the nominal compositions, candidate crystal structures, and experimental pattern features. The resulting reference phase combinations and mass fractions were used to benchmark MatDiffract across 40 valid experimental patterns.

Figure~\ref{fig:multiphase-identification}a presents the phase-combination identification accuracy for binary and ternary samples as a function of Top-k cutoff. Unlike single-phase identification, a multiphase identification result is considered correct only when all constituent phases in the candidate combination exactly match the reference set. Increasing the number of phases exponentially expands the combinatorial candidate space and amplifies uncertainty arising from peak overlap and low-abundance phases. For binary samples, initial MatDiffract retrieval achieved Top-1, Top-3, Top-5, and Top-10 accuracies of 75.0\%, 80.0\%, 90.0\%, and 95.0\%, respectively. Following full-pattern fitting and reranking, Top-1 accuracy improved to 85.0\%, Top-3 accuracy reached 90.0\%, and both Top-5 and Top-10 accuracies remained at 95.0\%. Thus, correct binary combinations are consistently ranked near the top of the candidate list and can be readily distinguished via subsequent full-pattern fitting.

Identification performance declined notably for ternary samples. The initial retrieval stage yielded Top-1, Top-3, Top-5, and Top-10 accuracies of 60.0\%, 70.0\%, 80.0\%, and 85.0\%, respectively. After full-pattern fitting, Top-1 accuracy increased to 70.0\%, with Top-3, Top-5, and Top-10 accuracies reaching 75.0\%, 85.0\%, and 90.0\%, respectively. The increased difficulty of ternary analysis does not stem merely from the addition of one more phase. With multiple phases contributing to the same diffraction pattern, the correct combination must be distinguished from a far larger set of competing candidate combinations. Low-abundance phases are especially susceptible to masking by intense major-phase peaks or background signals, which hinders them from ranking highly during initial retrieval. Nevertheless, correct ternary combinations typically appear within the Top-5 to Top-10 candidate range, providing a tractable candidate pool for downstream fitting and validation.

Figure~\ref{fig:multiphase-identification}b compares the Top-1 identification accuracy of JADE, initial MatDiffract retrieval, and MatDiffract with full-pattern fitting reranking. For binary samples, JADE reached an accuracy of 55.0\%, whereas MatDiffract reached 75.0\% prior to full-pattern fitting and 85.0\% after full-pattern fitting. For ternary samples, JADE performance dropped to 40.0\%, while MatDiffract maintained 60.0\% prior to full-pattern fitting and 70.0\% after full-pattern fitting. These results demonstrate that in multiphase systems with severe peak overlap and larger combination search spaces, MatDiffract outperforms conventional search-match workflows in retaining correct phase combinations, and uses full-pattern fitting to select candidates that provide a physically meaningful explanation of the overall experimental profile.

\begin{figure}[H]
\centering
\includegraphics[width=\linewidth,height=0.78\textheight,keepaspectratio]{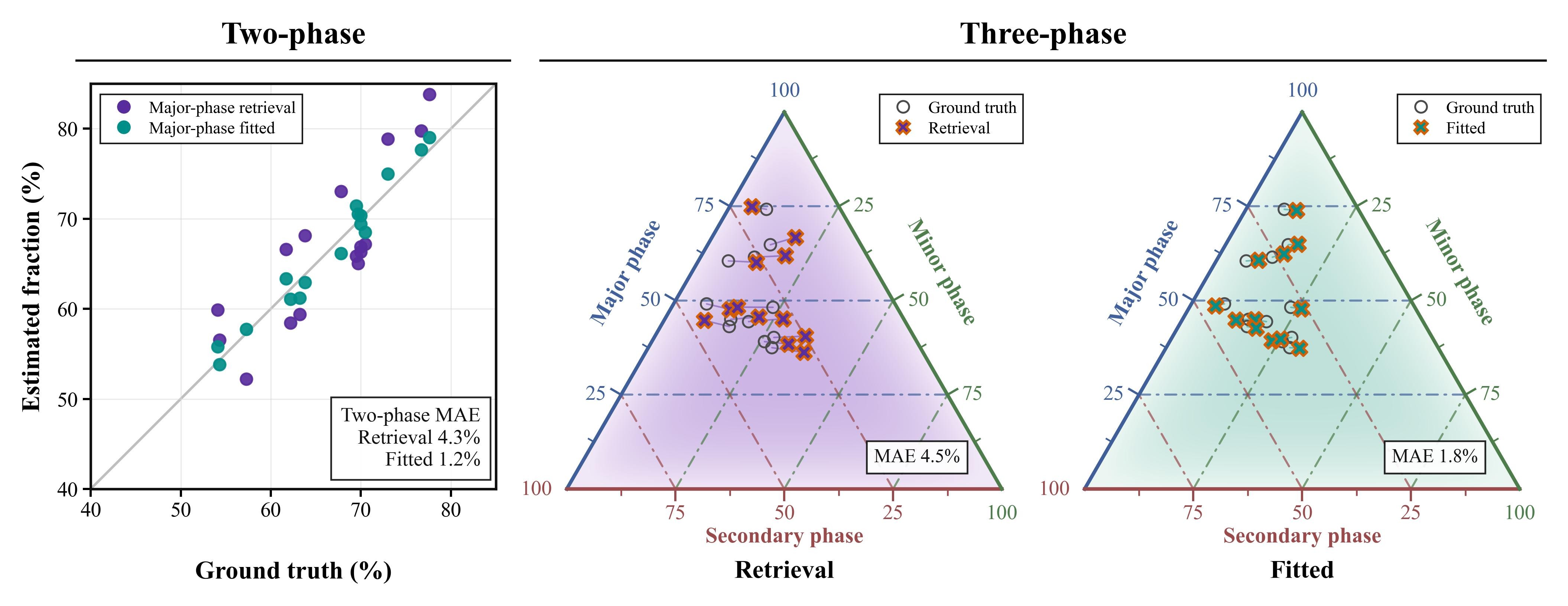}
\caption{Quantitative mass-fraction estimation for multiphase experimental XRPD patterns. Left: parity plot of major-phase mass fractions from initial retrieval and full-pattern fitting against reference values for binary samples. Middle and right: ternary composition diagrams for initial retrieval and full-pattern fitting results, respectively; open circles indicate reference compositions and crosses indicate estimated compositions. Corresponding mean absolute errors (MAEs) are labeled in each panel.}
\label{fig:mass-fraction}
\end{figure}

Beyond phase identification, multiphase XRPD analysis requires accurate quantification of phase mass fractions. Figure~\ref{fig:mass-fraction} demonstrates that full-pattern fitting substantially improved the accuracy of mass-fraction estimates. For binary samples, the mean absolute error (MAE) of the major-phase mass fraction decreased from 4.3\% at the initial retrieval stage to 1.2\% after full-pattern fitting, with data points converging toward the reference diagonal line. For ternary samples, the initially retrieved compositions exhibited noticeable deviation from the reference line, with an overall MAE of 4.5\%. Following full-pattern fitting, the estimated compositions moved closer to the reference line and the MAE was reduced to 1.8\%. Full-pattern fitting therefore enhances both candidate screening and the accurate partitioning of diffracted intensity among constituent phases.

The reduction in mass-fraction error arises from two complementary improvements. For binary samples, the primary challenge lies in partitioning intensity between the major and minor phases. The convergence toward the reference diagonal line after full-pattern fitting confirms that scale-factor and profile optimization corrects most of the initial fraction bias in fraction estimates. For ternary samples, errors are redistributed among the major, secondary, and minor phases, a trend more clearly visualized in ternary composition plots. While fitted compositions shift toward the reference line, low-abundance or weakly diffracting phases remain more sensitive to peak overlap, background interference, and masking by intense major-phase peaks. Accordingly, the results of Figure~\ref{fig:mass-fraction} should be interpreted as a substantial improvement from candidate combinations to quantitative compositions, rather than complete elimination of uncertainty for low-abundance phases.

\begin{figure}[H]
\centering
\includegraphics[width=\linewidth,height=0.78\textheight,keepaspectratio]{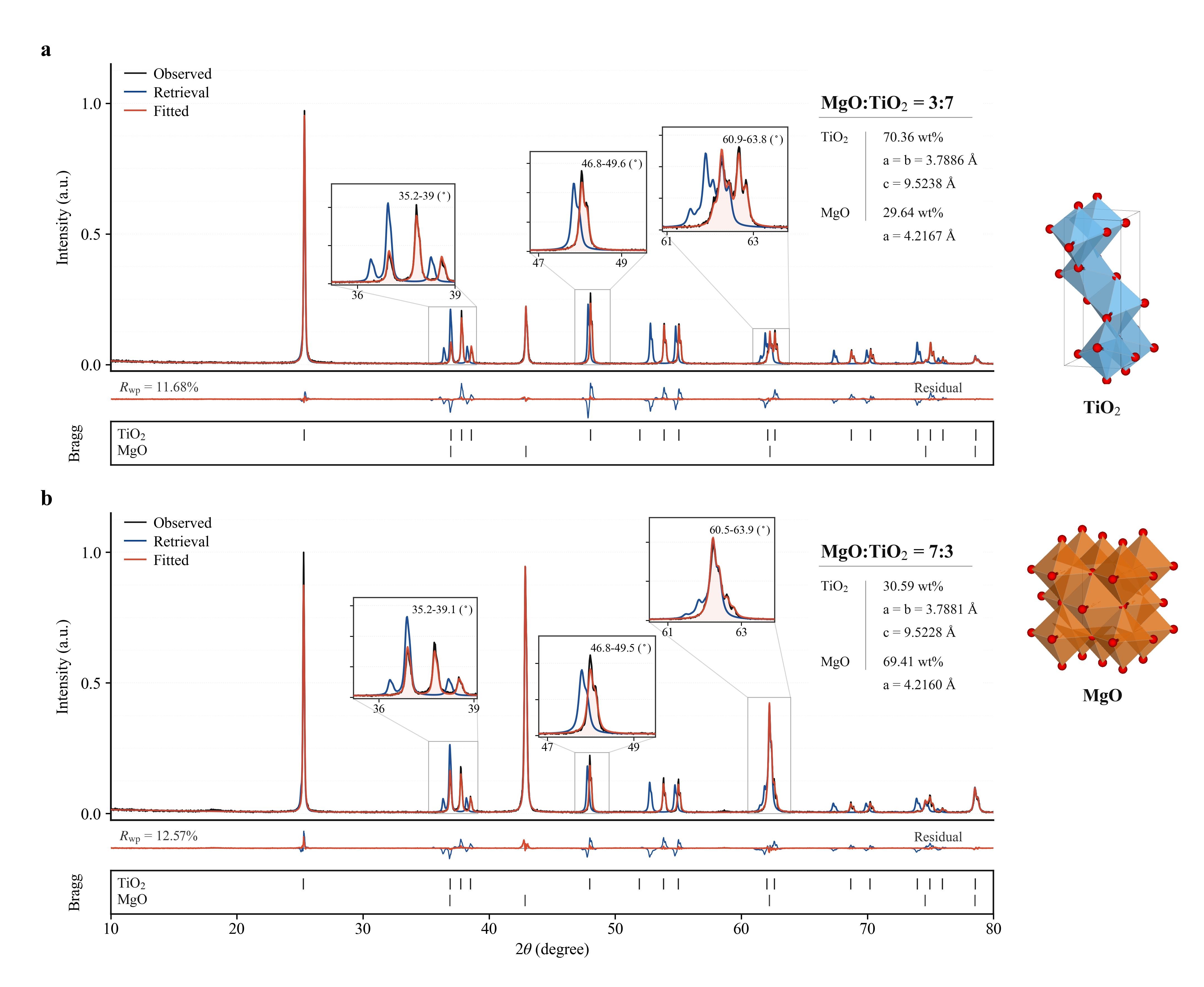}
\caption{Representative full-pattern fitting results, local peak alignment, and refined structural parameters for MgO--TiO₂ binary samples. (a) MgO:TiO₂ = 3:7 (Mg3-Ti7). (b) MgO:TiO₂ = 7:3 (Mg7-Ti3). Black, blue, and red curves denote the experimental pattern, the initial-retrieval calculation based on the unrefined TiO₂ unit cell, and the final fitted profile, respectively. Fitting residuals and Bragg positions for TiO₂ and MgO are plotted below each pattern. Insets highlight selected local 2θ regions showing peak-position and profile differences; right panels summarize fitted mass fractions, lattice parameters, and crystal structural models.}
\label{fig:binary-fitting}
\end{figure}

Figure~\ref{fig:binary-fitting} presents a comparative analysis of two MgO-TiO₂ samples with inverted nominal compositions: Mg3-Ti7 (MgO:TiO₂=3:7) and Mg7-Ti3 (MgO:TiO₂=7:3). For sample Mg3-Ti7, MatDiffract refined the TiO₂ and MgO mass fractions to 70.36 wt\% and 29.64 wt\%, respectively. The refined lattice parameters were a=b=3.7886Å and c=9.5238Å for TiO₂, and a=4.2167Å for MgO, with Rwp=11.68\%. For sample Mg7-Ti3, the fitted TiO₂ and MgO mass fractions refined by MatDiffract were 30.59 wt\% and 69.41 wt\%, respectively. The refined lattice parameters were a=b=3.7881 Å and c=9.5228 Å for TiO₂, and a=4.2160 Å for MgO, with Rwp=12.57\%. In both samples, the fitted fractions were in good agreement with the nominal ratios. In the TiO₂/MgO peak-rich windows of 35°-39°, 46.8°-49.6°, and about 60.5°-63.9°, the red fitted profile reproduces the experimental peak positions and relative intensities significantly better than the blue initial-retrieval profile calculated from unrefined crystal structures.

This comparison reveals that multiphase quantification error does not only originate from misallocation of phase fractions between TiO₂ and MgO. The as-calculated TiO₂ structure from the source database exhibits lattice parameters of a=3.80313Å and c=9.74365Å, which are systematically larger than the effective TiO₂ lattice parameters refined for both samples (a≈ 3.788Å and c≈9.523Å). According to Bragg\textquotesingle s law, an expanded unit cell increases interplanar d-spacings and shifts calculated diffraction peaks toward lower 2θ angles. Consequently, the systematic low-angle offset of the blue reference profile near TiO₂ peaks cannot be eliminated by merely rescaling the fractions of TiO₂ and MgO. Previous computational studies have established that DFT-calculated TiO₂ lattice parameters are sensitive to the choice of the exchange-correlation functional and computational parameters, with generalized-gradient approximation (GGA) calculations using the Perdew-Burke-Ernzerhof (PBE) functional often yielding appreciable deviations from experimental structures\supercite{labat2007}. By simultaneously refining lattice parameters, profile shape, and scale factors, MatDiffract corrects TiO₂ Bragg peak positions to match the experimental profile, while preserving phase fractions consistent with the nominal compositions.

Collectively, Figures~\ref{fig:multiphase-identification}-\ref{fig:binary-fitting} show three interconnected strengths of the MatDiffract workflow for multiphase analysis. First, it enhances phase-combination identification accuracy for both binary and ternary samples, with a more pronounced performance advantage over JADE observed for ternary mixtures. Second, the Top-k results confirm that correct combinations are consistently retained among the highest-ranked candidates, enabling users to focus subsequent full-pattern fitting and manual inspection on a compact candidate set. Third, as shown in Figures~\ref{fig:mass-fraction} and \ref{fig:binary-fitting}, full-pattern fitting not only validates phase combinations and optimizes mass fractions but also identifies and corrects systematic lattice-parameter deviations in database-derived crystal structures. The MatDiffract workflow therefore progresses from qualitative phase recognition to quantitative full-interpretation explanation and structural-model refinement.

Overall, MatDiffract delivers robust performance in phase-combination identification and mass-fraction estimation for binary and ternary experimental XRPD patterns. The challenges of peak overlap and low-abundance phase detection render multiphase analysis inherently more difficult than single-phase analysis, as reflected in the lower accuracy values for ternary samples. Nevertheless, MatDiffract achieved post-fitting Top-1 accuracies of 85.0\% for binary samples and 70.0\% for ternary samples, and reduced the mass-fraction MAEs to 1.2\% and 1.8\% for binary and ternary samples, respectively. These results confirm that the platform extends its capabilities beyond single-phase structure identification to multiphase mixtures routinely encountered in practical materials characterization.

\subsection{Platform Performance}\label{sec:performance}

The online analysis performance of MatDiffract was evaluated following construction of the full candidate database. Total analysis runtime encompasses the complete workflow from experimental-pattern input to final result output, and the dependence of retrieval time on candidate-database size was also taken into consideration. MatDiffract adopts a decoupled architecture comprising offline database construction and online pattern analysis. The offline workflow, including crystal structure calculation and simulated XRPD pattern generation, can be completed prior to analysis and expanded incrementally. The online workflow is dominated by experimental pattern preprocessing, vector search for candidate retrieval, and full-pattern fitting for structural refinement or mass-fraction optimization. This decoupled design eliminates redundant recalculation of simulated patterns for each experimental sample, such that online runtime is governed primarily by retrieval and subsequent refinement stages.

\begin{figure}[H]
\centering
\includegraphics[width=\linewidth,height=0.78\textheight,keepaspectratio]{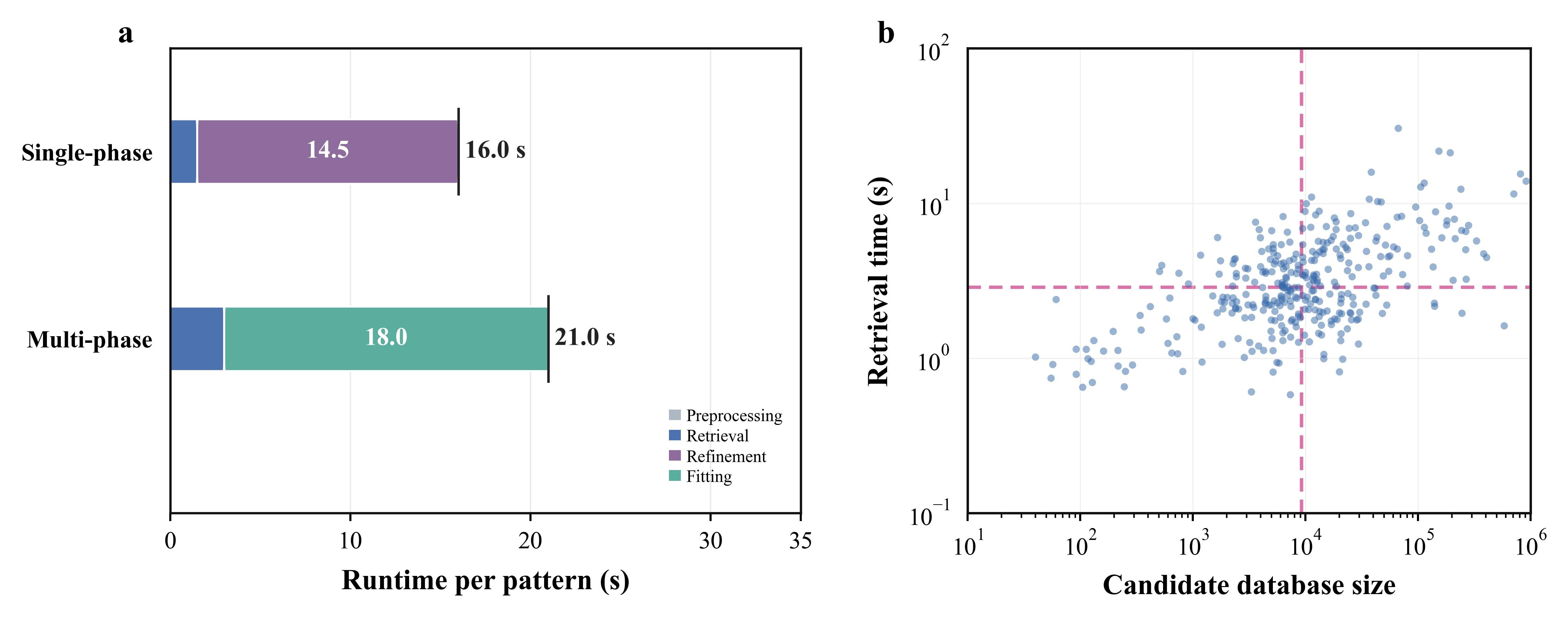}
\caption{Online analysis performance of MatDiffract and the effect of candidate-database size. (a) Runtime components for single-phase and multiphase samples after construction of the candidate database, including preprocessing, retrieval, structural refinement, and mass-fraction fitting. (b) Relation between candidate-database size and retrieval time. Points indicate validation samples; pink dashed lines denote median candidate-database size and median retrieval time. Offline candidate-database construction is not included in the online runtime.}
\label{fig:runtime}
\end{figure}

Figure~\ref{fig:runtime}a decomposes the online runtime into constituent stages for single-phase and multiphase samples. The single-phase workflow comprises pattern preprocessing, candidate retrieval, and structural refinement, yielding a total runtime of about 16.0s per pattern, with structural refinement accounting for the majority (\textasciitilde14.5s) of the computational cost. The multiphase workflow comprises pattern preprocessing, candidate-combination retrieval, and mass-fraction fitting, yielding a total runtime of about 21.0s per pattern, with mass-fraction fitting accounting for the majority (\textasciitilde18.0s). Collectively, these results demonstrate that MatDiffract completes online analysis of a single experimental XRPD pattern within tens of seconds once the candidate database is preconstructed.

The computational cost of the retrieval stage was analyzed by plotting retrieval time against candidate-database size on log-log axes (Figure~\ref{fig:runtime}b). The median candidate-database size was approximately 9.2×10³, with a corresponding median retrieval time of about 2.9 s. Larger candidate databases generally incur higher retrieval cost, but most samples remain within the sub-10-second range, with only a small number of exceptionally large candidate sets requiring tens of seconds or longer. Vector-database retrieval therefore efficiently narrows a large library of simulated-patterns into a rapidly rankable candidate pool, such that database expansion does not translate directly into a linear increase in manual screening effort. Given that structural refinement for single-phase samples and mass-fraction fitting for multiphase samples dominate the current runtime, retrieval is not the primary performance bottleneck of the current implementation. Future implementations of MatDiffract may reduce total runtime by restricting refinement to only a small number of high-ranked Top-k candidates. All runtime benchmarks were performed on the same workstation equipped with dual Intel Xeon Gold 6530 CPUs, 64 physical cores, 128 threads, 256GB memory, about 251GB available memory, and Python 3.9.23.

\section{Conclusions}\label{sec:conclusions}

In summary, we present MatDiffract, a material-informed computational platform for automated, high-throughput powder X-ray diffraction analysis designed to close the characterization throughput gap in modern materials discovery. Built on a first-principles inorganic crystal structure database, the framework integrates a perturbation-augmented simulated pattern database, multi-feature vector retrieval, and full-pattern fitting Rietveld refinement into a single end-to-end workflow. Unlike label-only machine learning classifiers, this architecture preserves rigorous crystallographic interpretability and physical consistency throughout the analysis pipeline, while retaining the speed and scalability of data-driven retrieval methods.

Systematic benchmarking on experimental single-phase and multiphase datasets demonstrates robust quantitative performance: 91.3\% Top-1 identification accuracy for single-phase samples, and 85.0\%/70.0\% Top-1 accuracy for binary/ternary mixtures, with mass fraction mean absolute errors below 2\%. All analyses are completed within tens of seconds per sample, matching the throughput requirements of automated synthesis platforms and high-throughput materials screening workflows.

A defining feature of the vector-based retrieval architecture is its native support for incremental expansion: new crystal structures, chemical systems, and perturbation conditions can be incorporated into the database without model retraining or label redefinition. This modularity distinguishes MatDiffract from fixed-dataset machine learning models, and enables seamless adaptation to emerging material systems and custom research needs.

The current validation focuses on well-crystallized multiphase mixtures. Moving forward, extending the framework to handle more complex systems, including solid solutions, amorphous backgrounds, strong preferred orientation, and \emph{in situ/operando} diffraction data, will further broaden its applicability. Integrating MatDiffract with self-driving laboratory workflows to enable closed-loop, feedback-driven materials synthesis and optimization represents a particularly promising direction. The underlying vector-retrieval methodology is also readily generalizable to other diffraction modalities, most notably neutron powder diffraction, opening avenues for cross-modal structural analysis.

Overall, MatDiffract provides a robust, extensible, and physically consistent solution for automated diffraction analysis, removing a key bottleneck in high-throughput materials characterization and accelerating the transition toward fully autonomous materials discovery workflows.

\section{Code Availability}\label{sec:code}

An online instance of the MatDiffract analytical workflow is hosted by the National High Energy Physics Scientific Data Center (NHEPSDC) and is accessible at \url{https://matdiffract.nhepsdc.cn}.

\section{Acknowledgments}\label{sec:acknowledgments}

This work was supported by the National Key R\&D Program of China (Grant No. 2024YFF0508500) and the open research fund of Songshan Lake Materials Laboratory (No. 2023SLABFK08). We are grateful to Prof. Erxi Feng for his insightful discussions and valuable guidance on the development, design, and result interpretation of MatDiffract. We thank the National High Energy Physics Science Data Center (NHEPSDC) for providing support for the deployment, operation, and online hosting of MatDiffract. We also acknowledge Dongguan Institute of Materials Science and Technology of the Chinese Academy of Sciences (DIMS) and Songshan Lake Materials Laboratory for their support in platform testing, data acquisition, and application validation.

\printbibliography[title={References}]

\end{document}